# Watt-level ultrahigh-OSNR single-longitudinal-mode tunable Brillouin fiber laser

Gaomeng Wang, Li Zhan*, Jinmei Liu, Tao Zhang, Jun Li, Liang Zhang, Junsong Peng, and Lilin Yi

*Department of Physics, State Key Lab of Advanced Optical Communication Systems and Networks, Key Laboratory for Laser Plasmas (Ministry of Education), Shanghai Jiao Tong University, Shanghai 200240, China*

*Corresponding author: lizhan@sjtu.edu.cn*

A watt-level ultrahigh optical signal-to-noise ratio (OSNR) single-longitudinal-mode (SLM) tunable Brillouin fiber laser (BFL) has been demonstrated. By optimizing the length of the single mode fiber (SMF) cavity at 11m and its output ratio at sixty percent, 1.04 W output power, as well as stable SLM operation is obtained at 2.24 W pump power. The single pass cavity BFL has the advantage that Brillouin pump frequency doesn't need to match the cavity mode, thus the stability is greatly improved. As only SMF is used in the cavity, the operate wavelength can be tunable without the restriction from self lasing cavity mode. Furthermore, it proves that core-pumped single frequency fiber laser is able to generate watt-level power. The laser has excellent performance in terms of noise, linewidth, and stability. © 2012 Optical Society of America

High power single-longitudinal-mode (SLM) fiber lasers with high optical signal-to-noise ratio (OSNR) have enormous applications such as remote sensing, ladar, coherent detection, and nonlinear optics. Recently, the output power of fiber lasers grew quickly, but SLM fiber lasers did not [1]. SLM fiber lasers can be boosted to watt-level [2,3], but it is still a challenge to preserve high OSNR. Besides, the output power of SLM fiber laser is also limited by the backward stimulated Brillouin scattering (SBS) [3,4].

Owing to narrow band gain of SBS, Brillouin fiber lasers (BFLs) with ~10m cavity can operate at SLM status. Though the threshold of BFLs [5] can be reduced to submilliwatt level, but the pump frequency has to match the cavity mode. This causes a big trouble for practical operation owing to the inevitable fluctuation of pump frequency. Moreover, it is difficult to output large power as the cavity is critically coupled [5]. Hybrid BFLs including active media don't need the critically coupled resonator [6], and they are able to achieve large power. A 1 W SLM Brillouin/ytterbium fiber laser with 60 dB OSNR has been demonstrated [7]. However, for a hybrid BFL, the wavelength tunability is limited at the self-lasing range [8]. Besides, it inevitably introduces amplified spontaneous emission (ASE) noise from active fibers. Different than hybrid BFLs, the BFLs without active fibers have wider tunable range, higher OSNR, and narrower linewidth. A SLM BFL with 12 m length cavity has been reported and shown stable output power [9]. In this letter, only SMF is used in the single pass cavity. By exploiting the high efficiency of SBS in fibers and the large output ratio of the lasing cavity, we demonstrate a watt-level SLM tunable BFL with 75dB OSNR.

Figure1 shows the configuration of the BFL, which consists of a tunable laser source (TLS) with 100 kHz linewidth, an erbium-doped fiber amplifier (EDFA), a polarization controller (PC), an optical circulator (OC) with 20% power loss, a 60/40 coupler with 5% loss. The cavity length is 11 m, corresponding to 18 MHz free spectral range. The laser from TLS is amplified by the EDFA as the input Brillouin pump (BP) which is injected to the single pass cavity clockwise. The threshold of this BFL is higher than the resonant Brillouin ring laser because the intra-cavity pump power in the latter is enhanced at resonance [5,10]. Although this scheme performs higher threshold, BP doesn't need to match the cavity mode. Thus our BFL can be tunable with improved stability [9,11].

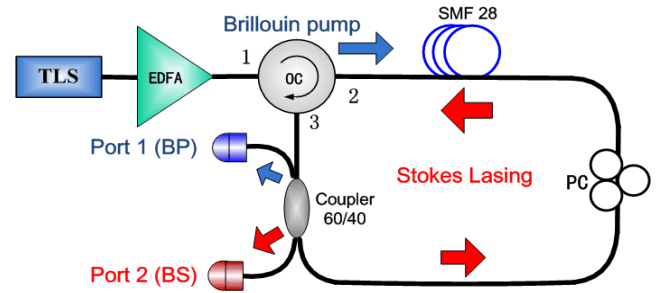

Fig. 1. Configuration of the proposed BFL.

Figure 2 shows the output power feature of the proposed laser at the wavelength of 1540 nm. The input threshold of this BFL is 920 mW as indicated by Fig. 2. Because the circulator has a 20% loss form port 1 to 2 or from port 2 to 3, the intra-cavity threshold is $920 \times 0.8 = 736$ mW. The output power reaches to 1.04 W at 2.24 W pump power. The transfer efficiency from BP to BS is 46%. The laser has a slope efficiency of 79% beyond the threshold, and it doesn't show any phenomena of saturation. The high slope efficiency attributes to the large output ratio in the cavity. PC is used to control the polarization states in the cavity and stabilize its output. We didn't adjust the PC before the output power from port 2 is less than 800 mW. The higher slope efficiency at high power region is obtained by adjusting the PC.

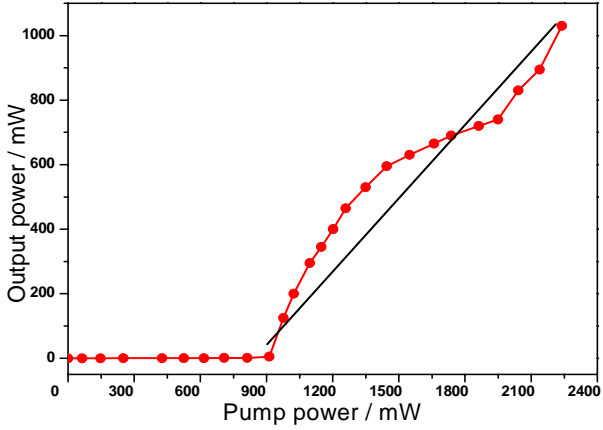

Fig. 2. BFL output power as the function of pump power from EDFA.

To estimate the threshold of the laser, we assume the pump and backward Stokes waves are polarized along the same direction, and the pump is resonant with a cavity mode at first. Thus, the threshold of ring cavity $P_{cr}$ is given by $R_m \exp(g_B P_{cr} L_{eff} / A_{eff} - \alpha L)=1$ [11,12], where, the peak Brillouin gain $g_B = 5 \times 10^{-11}$ m/w, the effective mode area of fiber $A_{eff} = 85 \mu m^2$, the attenuation constant $\alpha=0.0461$/km, the fraction of Stokes power fed back after each trip $R_m \approx 0.4$, $L$ and $L_{eff}$ are the fiber length and effective fiber length, $L=11$ m. It should be noted that the circulator has 20% loss, and the intra-cavity Brillouin pump is the needed threshold of proposed laser. The introduced loss by the circulator causes $R'_m = 0.8 \times R_m = 0.32$. So, we obtain $P_{cr} \simeq 176.1$ mW. At resonance, the intra-cavity power $P_{intra}$ [10] is enhanced over the input power $P_{input}$ by $1-\gamma_0/1-\kappa_r$, where, $\gamma_0=0.05$ is the loss of the coupler, $\kappa_r = (1-\gamma_0)T_m e^{-2\alpha L}$ is the resonant coupling coefficient. $T_m e^{-2\alpha L}$ is the power transmission for a ring resonator, and it is 80% here because the fiber loss can be neglected. Thus, we can deduce that $P_{intra} \simeq 4.0 P_{input}$, and the threshold in the single pass cavity should be 704.4 mW. Hence, we can conclude the experimental value of threshold (736 mW) is on reasonable range. Still, we didn't consider whether the peak of Brillouin gain completely overlaps with the cavity mode [11] and the polarization effect [13]

Figure 3 shows the output spectra from port 1 and port 2. The OSNR of the BFL is up to 75 dB which is much higher than the ones of hybrid BFLs [6,7]. The ultrahigh OSNR is obtained owing to the counter propagate, intensity and noise reduction of the SBS [14]. The frequency shift is 0.088 nm at 0.015 nm resolution. We believe that it is hard to obtain such high OSNR in other kinds of watt-level SLM fiber lasers.

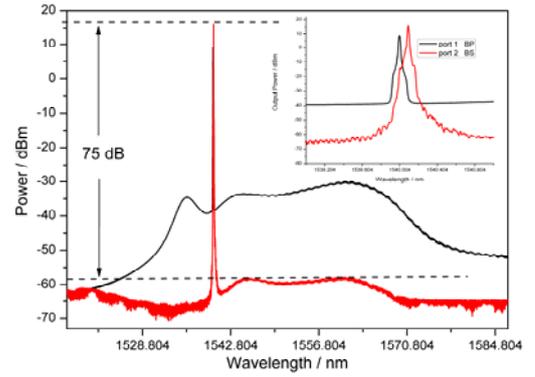

Fig. 3. The spectrum of BP and BS.

Because short cavity is adopted [7], the SBS gain profile provides enough discrimination to make the laser work on the SLM operation. The estimated linewidth of this BFL should be several kilo-hertz (kHz) [15]. Based on the large output power of this BFL, there is another good way to measure its linewidth [16]. The laser could be used as BP to generate the next order of Stokes in another fiber cavity, and then beat them together to generate the beating signal. The width of the beating signal should be wider than the linewidth of BS.

In Fig. 4(a), the observed beating frequency is 10.737 GHz when the BP is at 1560 nm, and its linewidth is 6 kHz as indicated by Fig. 4(b). Due to the narrowing effect of Stokes lasing, the second order of Stokes should be much narrower than the first one since a 10/90 coupler is used in the second cavity [17]. So the linewidth of this BFL should be less than 6 kHz. In the 50 MHz sweeping range in Fig. 4(a), only one line exists even at 1 W output power. It indicates that the SLM operation is achieved all times. No mode hopping is observed because the cavity is so short that only one cavity mode could exists in the 20 MHz Brillouin gain spectrum.

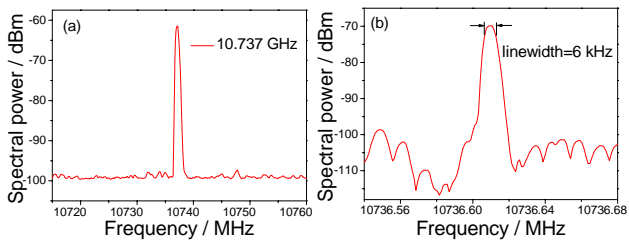

Fig.4. (a) The beating signal with 50 MHz sweeping range; (b) The beating signal with 200 kHz sweeping range.

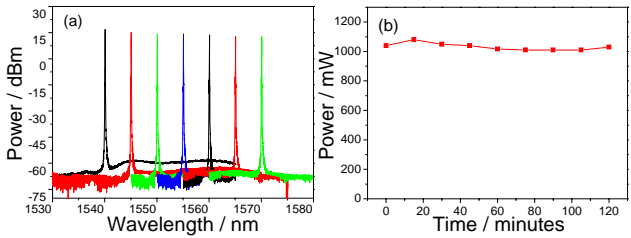

Fig.5. (a) The tunability of the proposed BFL; (b) Power fluctuation in 2 hours at 1 W output power.

For a laser, the broadband tunability will largely expand its practical value. Fig. 5(a) shows that the laser can be continually tuned in the range from 1540 nm to 1570 nm by adjusting the wavelength of TLS, because no gain fiber is used in the cavity. The operating range is only limited by the gain bandwidth of EDFA. The power stability at 1 W output is measured in 2 hours at the wavelength of 1540 nm. As shown in Fig. 5(b), the BFL output power fluctuation is below 4%. The stable operation is coincidence with the experimental result in ref. [9]. Due to the short cavity, only one cavity mode could exist in the narrow Brillouin gain spectrum. The large resonator loss (60% output ratio) also helps to the SLM lasing in the 20 MHz Brillouin gain band as other cavity modes can't obtain net gain. Consequently, a stable SLM BFL with 1W output power is achieved in the single pass SMF cavity. Obviously, our method to generate watt-level SLM fiber laser is economic and easy to operate. Aslo, it changes the usual view that the core-pumped single frequency fiber laser is impossible to generate watt-level output power [2]. Certainly, the laser would have lower threshold and more stable output if using polarization-maintaining fibers [13]. Such lasers can be found many applications, but also can provides a technological platform for superluminal propagation [18] or microwave generation [19].

In conclusion, a watt-level ultrahigh-OSNR SLM tunable BFL is demonstrated. Because of the high efficiency of SBS in SMF and large output ratio of the cavity, the output power reaches 1.04 W at 2.24 W pump power. At the same time, SLM operation is achieved. Ultra-high OSNR (75dB) is obtained which is an outstanding feature of proposed laser. Different from the hybrid BFL, this laser has broadband continually tunable range from 1540 to 1570 nm and ultrahigh OSNR. The measured linewidth is less than 6 kHz.

The authors appreciate J. K. Jang and S. G. Murdoch for helpful discussions and T. E. Murphy for helpful comments. This work is supported by the National Natural Science Foundation of China (NSFC) under grants 61178014 and 11274231, and the key project of the Ministry of Education of China under grant 109061.